\documentclass[aps,twocolumn,showpacs,showkeys,nofootinbib]{revtex4}
\usepackage{epsfig}
\usepackage{amsmath}
\usepackage{amsfonts}
\usepackage{amssymb}
\usepackage{graphicx}
\usepackage{colordvi}
\usepackage[T1]{fontenc}
\begin{document}

\title{Numerical Analysis of Galactic Rotation Curves}

\author{G. Scelza\footnote{e -
mail address: lucasce73@gmail.com}, A. Stabile$^1$\footnote{e -
mail address: arturo.stabile@gmail.com}}

\affiliation{$^1$Dipartimento di Ingegneria, Universit\`{a} del Sannio,
Palazzo Dell'Aquila Bosco Lucarelli, Corso Garibaldi, 107 - 82100,
Benevento, Italy}

\begin{abstract}
In this paper we present the discussion on the salient points of the computational analysis that are at the basis of the paper \emph{Rotation curves of galaxies by fourth order gravity} \citep{StSc}. The computational and data analysis have been made with the software Mathematica$^\circledR$ and presented at Mathematica Italy 5th User Group meeting (2011, Turin - Italy).
\end{abstract}

\keywords{Galactic rotation curves, numerical analysis.}
\maketitle

\section{Introduction}

The computational analysis here described is referred to the study of the galactic rotation curve. The theoretical details of the model investigated are omitted here, but fully available on the cited paper \citep{StSc}. The formula under study is $v(r,R,z)=\sqrt{r\frac{\partial}{\partial r}\Phi(r,R,z)}$ where $\Phi(r,R,z)$ is the gravitational potential
\begin{widetext}
\begin{eqnarray}\label{potential}
&&\Phi(r,R,z)\,=\,\frac{4\pi G}{3}\,\biggl[\frac{1}{r}\int_0^\infty
dr'\,\rho_{bulge}(r')\,r'\,\biggl(3\,\frac{|r-r'|-r-r'}{2}
-\frac{e^{-\mu_1|r-r'|}-e^{-\mu_1(r+r')}}{2\,\mu_1}
+2\,\frac{e^{-\mu_2|r-r'|}-e^{-\mu_2(r+r')}
}{\mu_2}\biggr)\biggr]
\nonumber\\\nonumber\\&&
+\frac{4\pi G}{3}\,\biggl[\frac{1}{r}\int_0^{\Xi}
dr'\,\rho_{DM}(r')\,r'\,\biggl(3\,\frac{|r-r'|-r-r'}{2}
-\frac{e^{-\mu_1|r-r'|}-e^{-\mu_1(r+r')}}{2\,\mu_1}
+2\,\frac{e^{-\mu_2|r-r'|}-e^{-\mu_2(r+r')}
}{\mu_2}\biggr)\biggr]
\nonumber\\\nonumber\\&&
-2\,G\,\biggr\{\int_0^\infty
dR'\,\sigma_{disc}(R')\,R'\,\biggl(\frac{\mathfrak{K}(\frac{4RR'}{(R+R')^2+z^2})}{\sqrt{(R+R')^2+z^2}}
+\frac{\mathfrak{K}(\frac{-4RR'}{(R-R')^2+z^2})}{\sqrt{(R-R')^2+z^2}}\biggr)+\int_0^\infty
dR'\,\sigma_{disc}(R')\,R'\,
\\\nonumber\\&&
\times\int_0^{\pi} d\theta'\frac{1}{3\,\sqrt{(R+R')^2+z^2-4RR'\cos^2\theta'}}
\biggl[e^{-\mu_1\sqrt{(R+R')^2+z^2-4RR'\cos^2\theta'}}
-4\,e^{-\mu_2\sqrt{(R+R')^2+z^2-4RR'\cos^2\theta'}}\biggr]\biggr\}\nonumber
\end{eqnarray}
\end{widetext}
where $\mathfrak{K}(x)$ is the Elliptic function and $G$ is gravitational constant. We remember that in the potential (\ref{potential}) we can distinguish the contributions of the bulge, the disk and the (eventual) Dark Matter. $r$ is the radial coordinate in the spherical system, while $R$, $z$ are respectively the radial coordinate in the plane of disc and the distance from the plane then we have the geometric relation $r\,=\,\sqrt{R^2+z^2}$. The main item is the choice of models of matter distribution. The more simple model characterizing the shape of galaxy is the following

\begin{eqnarray}\label{density_3}
\left\{\begin{array}{ll}
\rho_{bulge}(r)\,=\,\frac{M_b}{2\,\pi\,{\xi_b}^{3-\gamma}\,\Gamma(\frac{3-\gamma}{2})}\frac{e^{-\frac{r^2}{{\xi_b}^2}}}{r^\gamma}\\\\
\sigma_{disk}(R)\,=\,\frac{M_d}{2\pi\,{\xi_d}^2}\,
e^{-\frac{R}{\xi_d}}\\\\
\rho_{DM}(r)\,=\,\frac{\alpha\,M_{DM}}{\pi\,(4-\pi){\xi_{DM}}^3}\,\frac{1}{1+\frac{r^2}{{\xi_{DM}}^2}}
\end{array}\right.
\end{eqnarray}
where $\Gamma(x)$ is the Gamma function, $0\,\leq\,\gamma\,<\,3$ is a free parameter and $0\,\leq\,\alpha\,<\,1$ is the ratio of Dark Matter inside the sphere with radius $\xi_{DM}$ with respect the total Dark Matter $M_{DM}$. Moreove the couples $\xi_b$, $M_b$ and $\xi_d$, $M_d$ are the radius and the mass of the bulge and the disc. The parameters $\mu_1$ and $\mu_2$ are the free parameters in the theory and only by fitting process can be fixed.
A sensible item is the choice of distance $\Xi$ on the which we are observing the rotation curve. In fact all models for the Dark Matter component are not limited and we need to cut the upper value of integration in (\ref{potential}).

A further distinction are the contributions to the potential coming from terms of General Relativity (GR) origin and terms of Forth Order Gravity (FOG) origin. Finally our aim is the numerical evaluation of the rotation curve in the galactic plane

\begin{eqnarray}\label{velocity}
v(R,R,0)=\sqrt{R\frac{\partial}{\partial R}\Phi(R,R,0)}
\end{eqnarray}
Our analysis is then organized as follows: in section II we investigate the contribution of these terms on the galactic rotation curve, in section III a data fit between our theoretical curves and the data of the rotation curve of the Milky Way and the galaxy NGC 3190 and in section IV we report the conclusions.

\section{The computation}

\begin{figure}[t]
  \centering
  \includegraphics[scale=0.55]{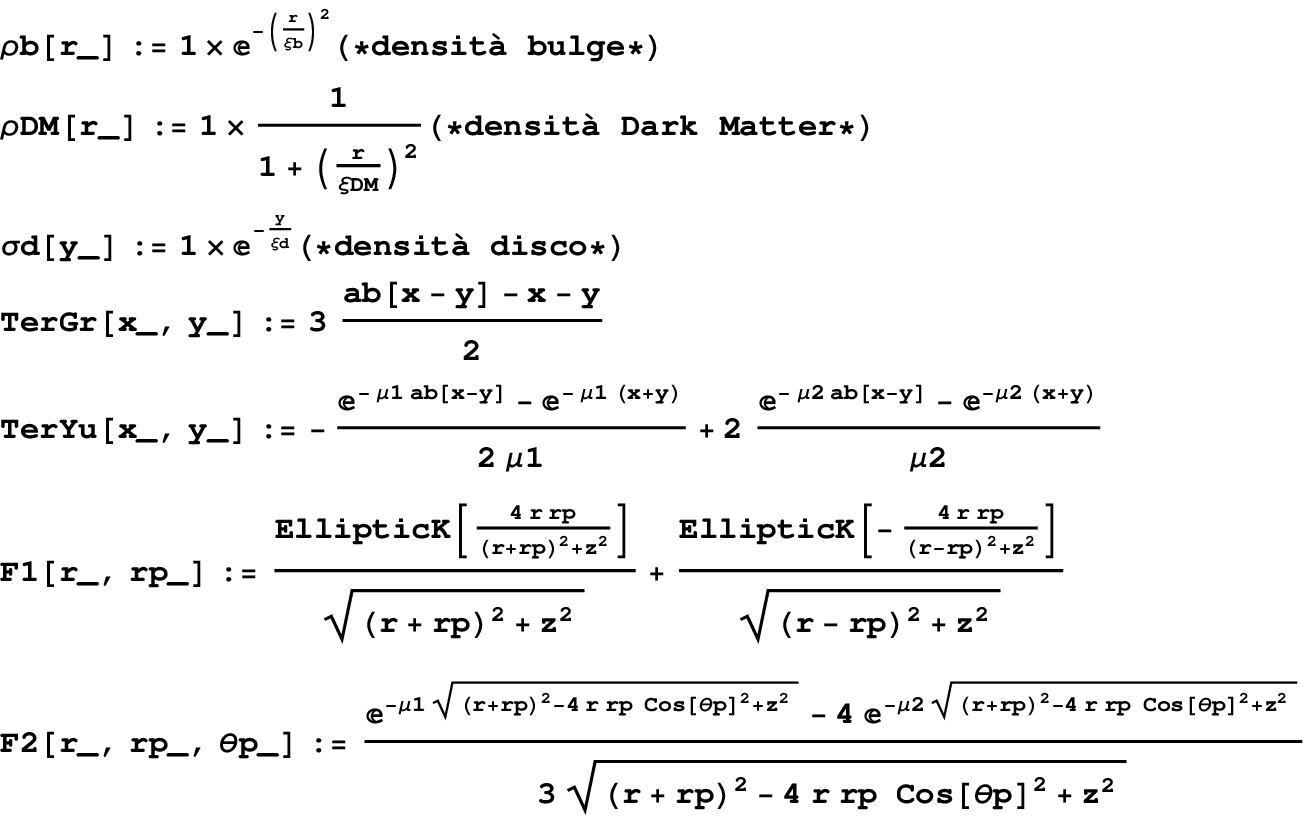}\\
  \caption{The definition of the density terms and the splitting of GR contributions and FOG contributions to the gravitational potential}
  \label{schermata_1}
\end{figure}

\begin{figure}[t]
  \centering
  \includegraphics[scale=0.55]{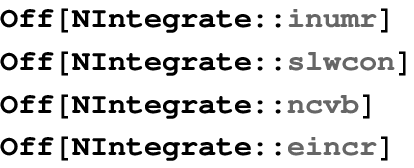}\\
  \caption{Shooting off the warning messages}
  \label{schermata_2}
\end{figure}

\begin{figure}[t]
  \centering
  \includegraphics[scale=0.55]{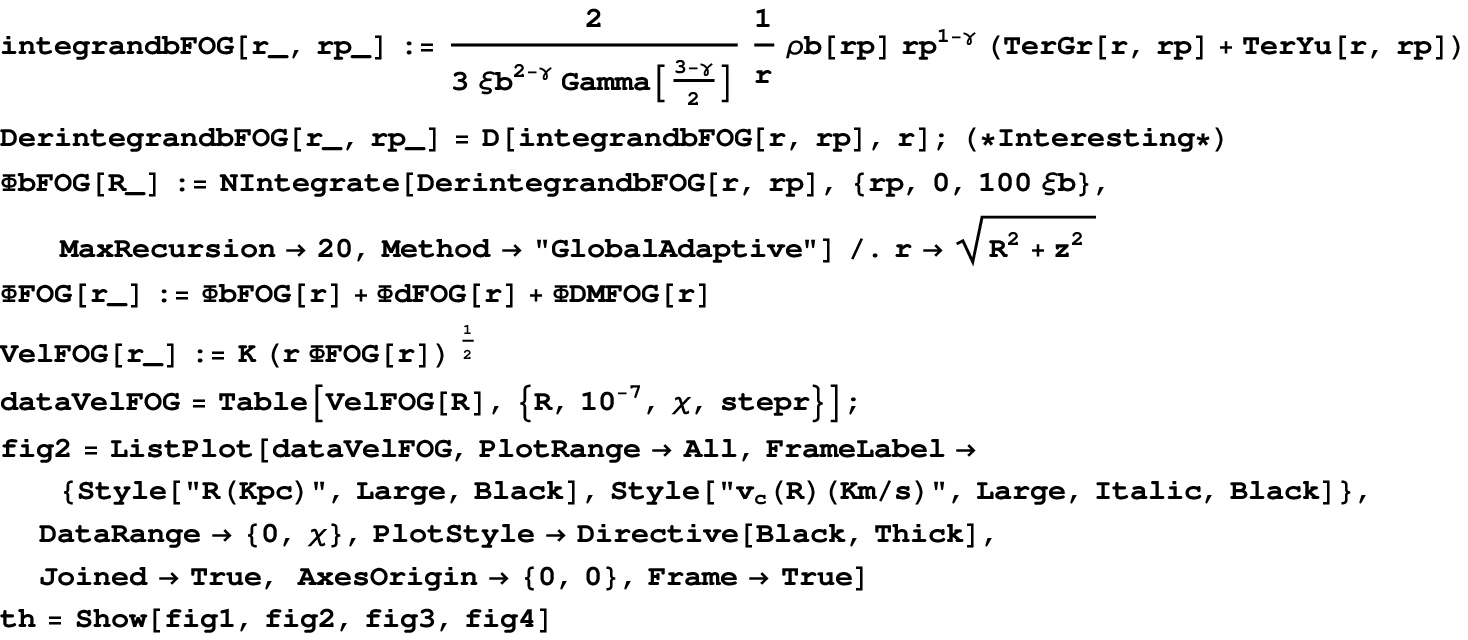}\\
  \caption{The derivative of the bulge term in the potential (case FOG).}
  \label{schermata_3}
\end{figure}

The first step, after the definition of the numerical values for the parameters, has been the building of the velocity starting from de derivative of the potential as we can see in figure \ref{schermata_1}. The derivative and integration operations commute, then we ``transport'' the derivative in the the integrand and then we make the integration. We found this computationally more rapid. Moreover, we make a splitting in the GR contributions and FOG contributions in the gravitational potential. Then it follows the turning off the warning messages concerning the numerical integrations as showed in figure \ref{schermata_2}. Indeed for the first \verb|Off|, as we can see in figure \ref{schermata_3}, all the definitions are made with the ``SetDelayed'' command that postpones the numerical evaluation of the integral making it not immediately numerical. The following warnings inform us of the need to increase the precision of the computation. An interesting thing to note in figure \ref{schermata_3}, is that in the definition of the derivative by means of mute variables, it need not a ``SetDelayed'' command, but a simple ``='' command.
  
\section{Data fit}

The next and more interesting step, is the comparison of the experimental data and what predicted by our model. From the literature cited in \citep{StSc} we can obtain the galactic speed values as function of the distance from the center and the corresponding errors. For instance, we show in some detail the manipulation of the data coming from the analysis of \citep{Stark}, concerning the external part of the Milky Way.

\begin{figure}[t]
  \centering
  \includegraphics[scale=0.55]{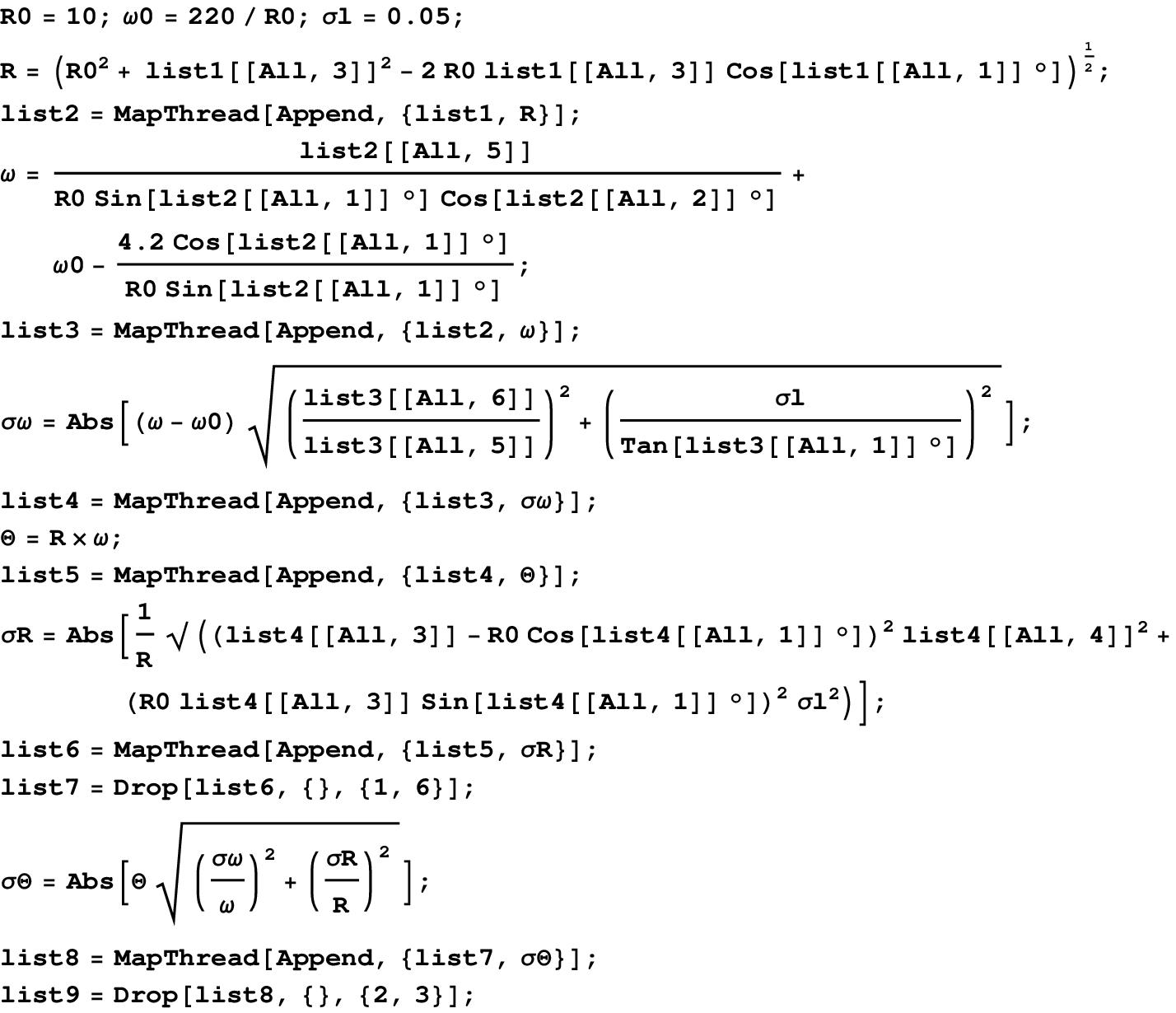}\\
  \caption{Experimental data manipulation.}
  \label{schermata_4}
\end{figure}

\begin{figure}[t]
  \centering
  \includegraphics[scale=0.55]{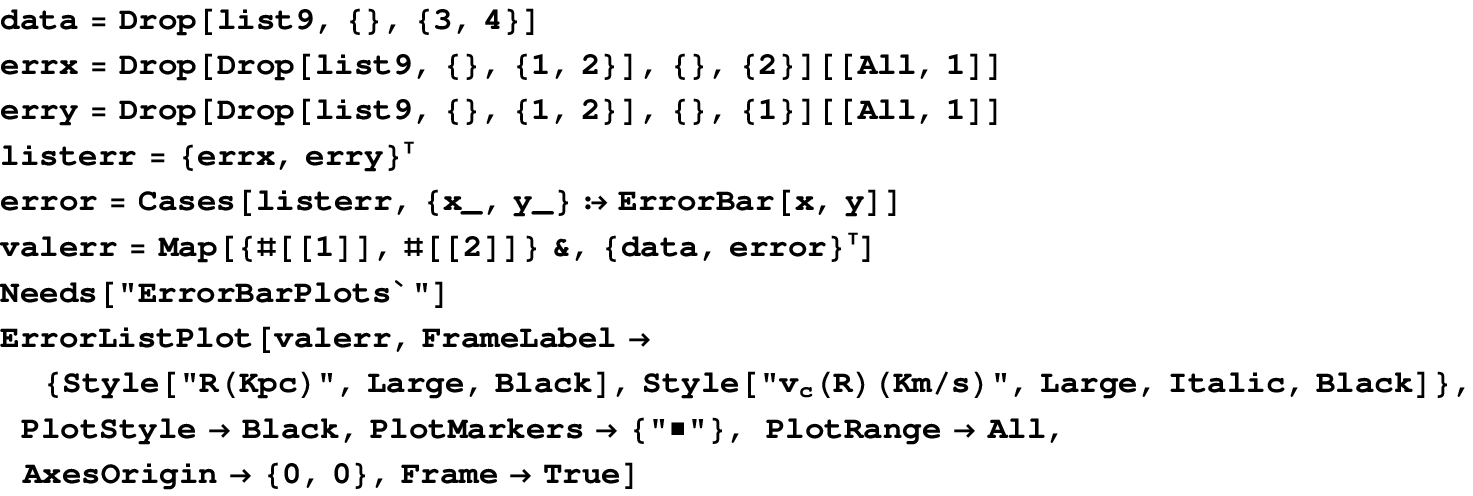}\\
  \caption{ErrorListPlot procedure.}
  \label{schermata_5}
\end{figure}

We start copying the data listed in the table 1 of \citep{Stark} in a table called \verb|list1|. Then we follow the prescriptions given by the authors with the introduction of new variables. As it is possible to see in figure \ref{schermata_4} we preserve the same notations and append to the initial \verb|list1| the new variables. For instance, for the $R$ variable, with the command \verb|MapThread[Append,{list1,R}]|, we obtain a new table, here \verb|list2| with one more column, the $R$'s valuer. And so on with the other variables. We computed, with the usual procedures, the errors on these derived quantities, here written as $\sigma x$. Then $\sigma$R and $\sigma\theta$ are, respectively, the error bars on the radius (the distance from the galactic center) and on the corresponding speeds and we process them together with the data so as shown in the figure \ref{schermata_5}.

\begin{figure}[t]
  \centering
  \includegraphics[scale=0.8]{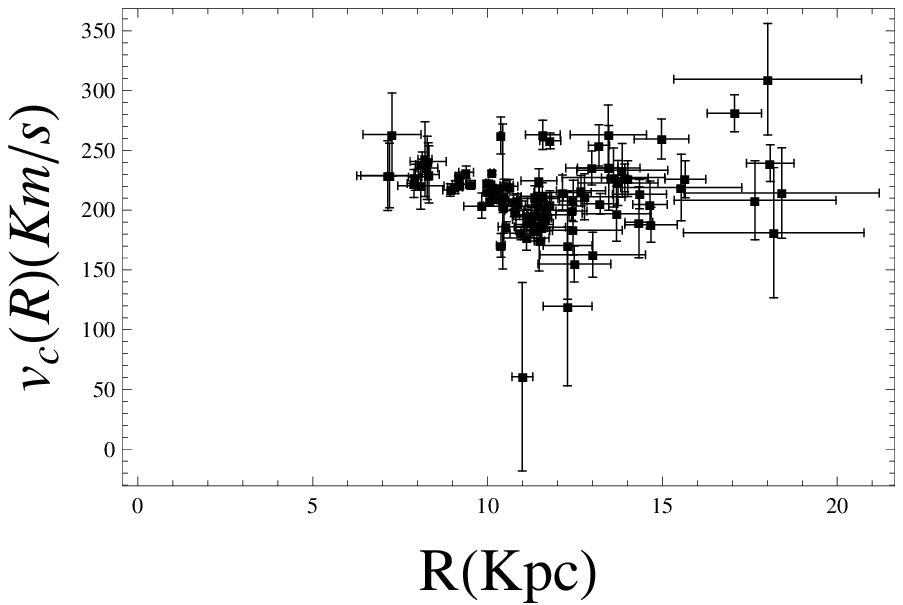}\\
  \caption{ErrorListPlot of the experimental data.}
  \label{Stark}
\end{figure}

In figure \ref{Stark} it is shown the result. With the command in the second line of figure \ref{schermata_5}, we obtain a list whose elements are of kind \verb|ErrorBar[err_x,err_y]|. In the third line we build a list whose element are \verb|{{x,y},ErrorBar[err_x,err_y]}|. In these conditions, we need to load the package \verb|ErrorBarPlot| in order to make an \verb|ErrorListPlot|. Similar procedures for the others two part of the Milky Way data and for the NGC 3190 data.

At this point we proceeded following two strategies.

The first one, the faster, has been to overlap the theoretical graphs with the experimental one using the command \verb|Show|. In this case, the values of parameters in the densities (bulge, Dark Matter and disk) and of reduced masses, $\mu_1$ and $\mu_2$, see screen shot in figure \ref{schermata_1}, are chosen by a direct overlap of the graphs.

The second strategy, more rigorous and slower, is the fit procedure. In this case, we fix all other parameters except the "masses''  $\mu_1$ and $\mu_2$. These variables are the values that must be found in the find fit procedure.

\begin{figure}[t]
  \centering
  \includegraphics[scale=0.55]{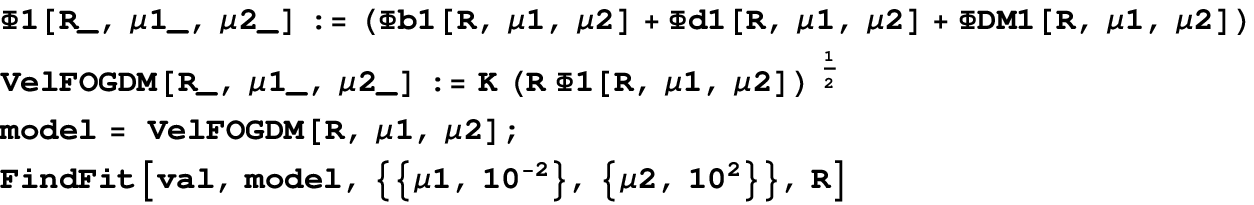}\\
  \caption{FindFit procedure. The ``masses'' $\mu_1$ and $\mu_2$ are found by the fit with the experimental data, here represented by $val$.}
  \label{schermata_8}
\end{figure}

We note that the \verb|FindFit| procedure uses the parameter constraints option. In this way, it is possible to eliminate all the solutions not physically allowed and to find the values obtained by the direct investigation, that is the first strategy, $\mu_1=10^{-2}\,a^{-1}$, $\mu_2=10^{2}1,a^{-1}$ where $a$ is the characteristic scale length fixed to the value of 1 Kpc.

\section{Conclusions}
\begin{figure}[t]
  \centering
  \includegraphics[scale=0.45]{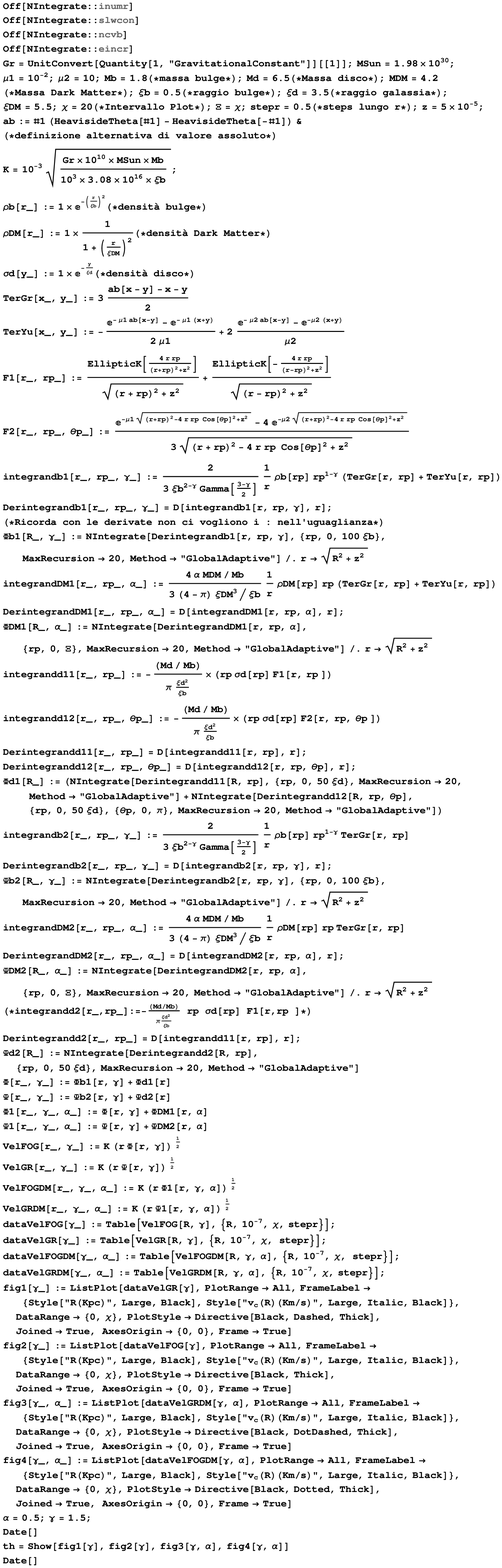}\\
  \caption{Screen-shot of the full program for the rotation curve of the Milky Way (figure \ref{plot_1_PRD}).}
  \label{schermata_9}
\end{figure}

\begin{figure}[t]
  \centering
  \includegraphics[scale=0.8]{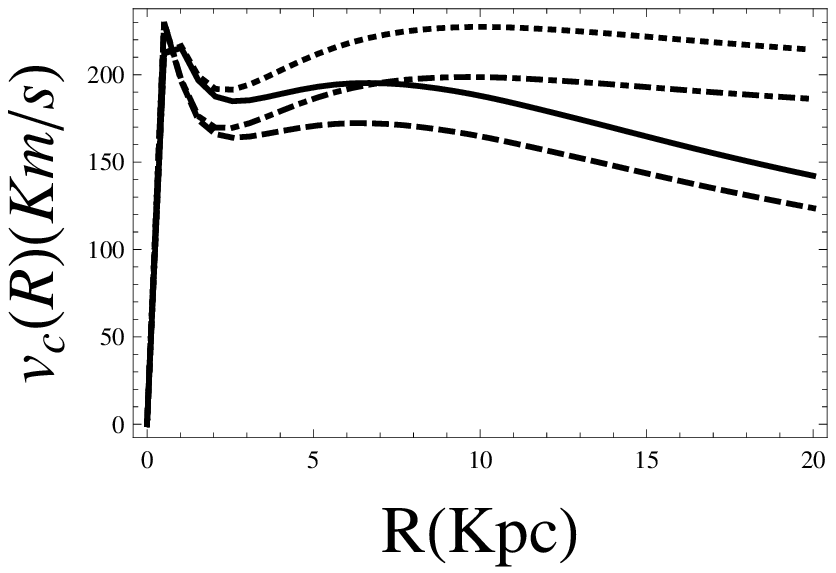}\\
  \caption{Plot of the galactic rotation curve by using the full program for Milky Way (figure \ref{schermata_9}). The cases are the following: GR (dashed line), GR$+$DM (dashed and dotted line), FOG (solid line), FOG$+$DM (dotted line). The values of masses are $\mu_1\,=\,10^{-2}\,\text{Kpc}^{-1}$ and $\mu_2\,=\,10^2\,\text{Kpc}^{-1}$ \citep{StSc}}
  \label{plot_1_PRD}
\end{figure}

In this paper we presented the salient points in the program we build in the computation of the velocity curves of the Milky Way and the galaxy NGC 3190. In figure \ref{schermata_9} is shown the full code corresponding to the plot of the figure \ref{plot_1_PRD} \citep{StSc}, that is the code for a galaxy whose components are the bulge, the disk and the Dark Matter. The code referring also to the study of the galaxy NGC 3190 is exactly the same with the exclusion of the part of code referring to the bulge.

\begin{figure}[t]
  \centering
  \includegraphics[scale=0.8]{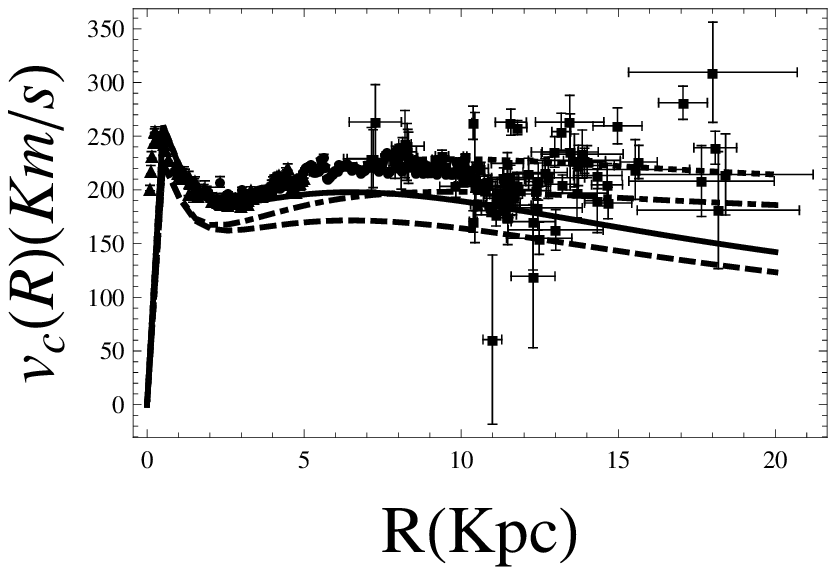}\\
  \caption{Superposition of theoretical behaviors GR (dashed line), GR$+$DM (dashed and dotted line), FOG (solid line), FOG$+$DM (dotted line) by using the full program (figure \ref{schermata_9}) on the experimental data for Milky Way. The values of masses are $\mu_1\,=\,10^{-2}\,\text{Kpc}^{-1}$ and $\mu_2\,=\,10^2\,\text{Kpc}^{-1}$ \citep{StSc}.}
  \label{plot_2_PRD}
\end{figure}

As it is possible to see from figure \ref{plot_2_PRD} \citep{StSc}, the agreement of our model with the experimental data of the Milky Way is very good. Only for very low values of the distance $R$ the agreement is not perfect. This suggest us that we only need an improvement of the parameters in the code, maintaining the code itself essentially unchanged.

\end{document}